# Threat or Opportunity? - Examining Social Bots in Social Media Crisis Communication

**Florian Brachten**
University of Duisburg-Essen
Duisburg, Germany
Email: florian.brachten@uni-due.de

**Milad Mirbabaie**
University of Duisburg-Essen
Duisburg, Germany
Email: milad.mirbabaie@uni-due.de

**Stefan Stieglitz**
University of Duisburg-Essen
Duisburg, Germany
Email: stefan.stieglitz@uni-due.de

**Olivia Berger**
University of Duisburg-Essen
Duisburg, Germany
Email: olivia.berger@stud.uni-due.de

**Sarah Bludau**
University of Duisburg-Essen
Duisburg, Germany
Email: sarah.bludau@stud.uni-due.de

**Kristina Schrickel**
University of Duisburg-Essen
Duisburg, Germany
Email: kristina.schrickel@stud.uni-due.de

## Abstract

Crisis situations are characterised by their sudden occurrence and an unclear information situation. In that context, social media platforms have become a highly utilised resource for collective information gathering to fill these gaps. However, there are indications that not only humans, but also social bots are active on these platforms during crisis situations. Although identifying the impact of social bots during extreme events seems to be a highly relevant topic, research remains sparse. To fill this research gap, we started a bigger project in analysing the influence of social bots during crisis situations. As a part of this project, we initially conducted a case study on the Manchester Bombing 2017 and analysed the social bot activity. Our results indicate that mainly benign bots are active during crisis situations. While the quantity of the bot accounts is rather low, their tweet activity indicates a high influence.

# 1    Introduction

Social media has altered the way in which crises (man-made or natural catastrophes) are perceived and communicated about. Social media has become a key instrument for information sharing that allows quick interactions due to short response times and is an efficient way to quickly reach a vast amount of users (Mirbabaie et al., 2014; Palen et al., 2010; Stieglitz et al., 2017b; Wagner et al., 2012). Simultaneously, social media crisis communication is a highly complex topic, because of the "*high number and dynamics of participants as well as the technical requirements of platforms*" (Stieglitz et al. 2017, p. 4). As crisis situations are oftentimes distinguished by their uncertainty through a lack of information, people try to reduce this uncertainty and close information gaps by communicating with each other and thus making sense of the situation (Dervin, 2003; Weick, 1988). According to sensemaking, individuals are living in a world of uncertainties and knowledge gaps (Dervin, 2003). Social media platforms could serve as a relevant medium to close these gaps as they allow their users to decrease their uncertainty through an information exchange process. Research revealed that social media platforms can be highly effective for all participants to make sense of an event (Mirbabaie and Zapatka, 2017). However, the characteristics of social media platforms at the same time give way to negative occurrences such as the spread of false information (Stieglitz et al., 2017b). One phenomenon that has recently been observed to foster these negative mechanics are so called *social bots* – accounts on social media platforms that mimic human behaviour and spread vast amounts of messages believed to aim at influencing human user's opinions. These accounts have seen a rise in attention lately, oftentimes in the context of a possible interference in politics. While the discussion about possible negative effects of social bots predominates the discussion, there are also cases of benign bots, that is bots which are applied in support of a certain topic (Brachten, et al. 2017).

Prior research indicates that social bots might also be active during crisis situations as well (Bunker et al., 2017). However, research on the occurrence of social bots during crisis communication on social media has been rather sparse. Especially as emergency agencies have started to be active on social media and seek to manage the communication (Ludwig et al., 2015), it is important to understand the validity of the data the agencies are working with. We seek to shed light on this phenomenon. Thus, we initially aim to analyse a crisis situation to elucidate the occurrence of social bots. To do so, we seek to present preliminary findings for the two following research questions:

> *RQ1:    To what extent are social bots active during a crisis situation?*
> *RQ2:    What content do influential social bots spread during a crisis situation?*

In order to answer these research questions, we conducted a case study on the Manchester Bombing 2017. By filtering the dataset for users with the highest retweet count, we identified the so-called power users, which have the biggest influence on crisis communication and therefore exert a high impact on sensemaking. The findings will serve as a first step of better understanding the value of social media communication during crisis situations. This research-in-progress is part of a bigger project that aims at identifying the impact of social bots on crisis communication to a full extent.

# 2    Background

## 2.1    Crisis Communication in Social Media

Social media has gained increased importance in crisis situations as an information source as well as a communication channel (Ehnis et al., 2014; Ross et al., 2018). Prior studies have shown that during crises and risk events, people engage in several forms of communication to learn about the specific situation, in order to gain control and reduce their personal uncertainties (Lachlan et al., 2016, 2009). Social media platforms, such as Facebook or Twitter initially gained importance, as they became a highly utilized resource for communication and information seeking during crisis events (Stieglitz et al., 2017b). Acar and Muraki (2011) examined crisis communication on Twitter during the tsunami in Japan in 2011 and revealed that people who were directly involved, tweeted about their unsafe situation and survival related topics while people from indirectly related areas tweeted to inform others that they were safe. Research also shows that especially for emergency agencies trying to manage a crisis and coordinate people affected by it, social media platforms have become an important way to reach people affected by a crisis situation. As described above, the reason for people to flock to social media in crisis situations is the need for information. As the need for disclosed information increases and the amount of new validated information stagnates, people tend to form their own opinion through dialogue with other users a process called sensemaking (Maitlis and Christianson, 2014).

## 2.2　The Theory of Sensemaking

Sensemaking has become an important topic in the study of organisations and IT systems (Maitlis and Christianson, 2014) and takes place in times of vague or missing information. Sensemaking "*is triggered by cues, such as issues, events or situations for which the meaning is ambiguous and outcomes are uncertain*" (Maitlis and Christianson 2014, p. 70), criteria, all of which apply to crisis situations. The main goals of sensemaking are the arrangement of information in a meaningful context and the reduction of individual knowledge gaps (Stieglitz et al., 2018). The general concept of sensemaking has been predominantly researched within the domains of organisations and on the individual level. However, social media enables users to participate in a collaborative sensemaking process, e.g. by reading, retweeting, and commenting on microblog sites like Twitter. Thus, social media hereby serves as a medium for collective sensemaking and research has shown that a loosely connected group of people can work together to come to an agreement about a certain situation (Hughes et al., 2008). The connectivity and fast information flow in social media can be a factor that further facilitates this collective sensemaking process. As sensemaking in social media is a collaborative process, each individual can influence the group's opinion (Stieglitz et al., 2018). This is an important aspect as the concept relies on the assumption that human users in crisis situations on social media have the goal to work together to reduce uncertainty and make sense of a situation. However, research indicates that a portion of communication on social media platforms does not stem from human users but is generated by automated actors, so called social bots. On Twitter for example, studies estimate that between 9% and 15% of the overall traffic are actually generated by bots (Varol et al., 2017).

## 2.3　Social Bots

The term "social bot" describes accounts on social media sites that are controlled by bots and imitate human users to a high degree but differ regarding their intent (Stieglitz et al., 2017a). In the last years this phenomenon has gained growing interest. While the term social bot can describe both – benign as well as malicious bots, especially the latter have seen a rise in attention. Especially in the political sphere, their potential influence on the opinion of social media users and thus the outcome of votes has been researched (e.g. in the presidential election 2016 by Bessi & Ferrara (2016)), where an activity of social bots could be proven. A key aspect of the bots is the imitation of human behaviour, for social media users it is oftentimes hard to tell whether a message on a platform stems from a bot or a human user. These bots can thus evoke the impression that some information or opinion, regardless of its accuracy, is highly popular and endorsed by many, a strategy oftentimes observed the literature on bots (Ferrara et al., 2014). As this example suggests, there have been cases where social bots interfered with the communication on social media. However, the findings on crisis communication remain sparse. While Cassa et al. (2013) did find bot communication in their dataset, it has not been the main research focus. Still, as social media gains more importance for emergency agencies to reach people and for the people affected by a crisis to make sense of it and gain information, it is important to get an understanding on the role social bots play in these scenarios. Should they for example try to spread false information, this has to be taken into account when assessing the usefulness of social media data.

# 3　Research Method

## 3.1　Data Collection

Twitter Data has been researched extensively in previous literature and serves as the main social media platform within the IS community. One of the reasons is the open mentality and the API. As previous findings indicate that the number of bots applied may be related to the language of an event with English providing a richer database (Brachten et al., 2017), we chose to examine a case that took place in an English speaking area. Due to medias' prominence which indicates a high Twitter activity, we selected the Manchester bombing. On May 22 in 2017 at 22.35 CEST, an individual detonated a bomb in the Manchester Arena. A hashtag was created within the social media communication (#roomformanchester). We collected data via the Twitter Search API for the hashtags "#roomformanchester", "#Manchester" and "#manchesterbombing", and the keyword "Manchester". The data consisted of tweets sent between 22/05/2017 at 22:00 CEST and 24/05/2017 at 22:00 CEST, as this was the period of time with the highest tweet activity.

## 3.2　Bot Detection

For an initial and representative analysis, we focused on the top 20,000 power users as those have the highest impact on sensemaking (Oh et al., 2015). We identified the bots within the sample by using the tool *Botometer*. The tool enables users to enter a Twitter screenname and then assigns a value between 0 and 1 indicating the probability of that account being a bot (Davis et al., 2016). This estimation is based on more than 1.000 features of an account (e.g. the tweet frequency or friend-follower-ratio) and has been applied in past studies for that purpose (Varol et al., 2017). Former studies applied a value of 0.7 and above for an account to be labelled as a bot (Bessi and Ferrara, 2016), which we follow to discriminate between bots and human accounts.

### 3.3 Content Analysis

Following this **first step**, a content analysis was conducted (Skalski et al., 2017). The tweets of the identified bots were classified into dedicated types two general categories from previous research - malicious or benign (Stieglitz et al., 2017a). To classify a bot as malicious, we classified the content of a tweet as (1) commercial, (2) misdirection, (3) spam or (4) others_malicious (Subrahmanian et al., 2016). To classify a bot as benign, we labelled the content as (1) news reports, (2) weather reports, (3) sports bots, (4) traffic bots or (5) others_benign (Lokot and Diakopoulos, 2016). As the application of the categories from the first step showed that all of those tweets fell in the category of 'benign', as a **second step** we conducted a second analysis by adapting a codebook from Skalski et al. (2017) to our case, which specifies additional subcategories.': (1) Help, (2) Succor, (3) Sympathy, (4) News Report, (5) Search, (6) Criticizing fake information and (7). Other. To ensure a complete understanding of the categories a test coding of 100 tweets was conducted by three coders. Afterwards, Krippendorff's alpha was calculated to test the intercoder reliability. The achieved value of 0.889 indicates a complete coding consistency.

## 4 Preliminary Findings

The initial dataset consisted of 3,285,906 tweets by 1,455,148 accounts, of which tweets from 63,965 accounts were retweeted. Table 1 provides an initial insight into the descriptive statistics for the dataset and shows a rather high average Tweet count (though with a high SD) for the accounts in the sample. To gain first insights into the data, we filtered the number of tweets for preliminary analysis. Therefore, the data was pre-processed manually to identify the 20,000 top users. These power users are those accounts that were most retweeted during the tracking period (Oh et al., 2015). They are participants who are central in a network and who are influencing other participants, hence might have an influence on the sensemaking process. This influence is reflected by their high number of retweets (Mirbabaie and Zapatka, 2017).

|        | Accounts in the sample | | Tweets in the sample | |
|--------|-------------|----------------|---------------|----------------|
| Metric | Tweet count | Follower count | Retweet count | Favourite count |
| Min    | 0           | 0              | 0             | 0              |
| Max    | 8,027,281   | 76,966,400     | 107,694       | 317,964        |
| Mean   | 30,052      | 4,493          | 5,679         | 2.18           |
| SD     | 67,513      | 130,385        | 13,589        | 245.85         |

*Table 1. Descriptive statistics for the initial dataset*

During the bot detection on the sample of the 20,000 power users 58 accounts (0.2%) were assigned a value greater than 0.7 and were thus classified as bots. These accounts posted a total of 405 tweets within the sample of 20,000 top-users (approximately *7 tweets per account*). Our content analysis revealed that all bot accounts fall into the general category of benign bots. Table 2 shows the distribution of tweets in the categories.

| Classification | Number of tweets |
|---|---|
| Search | 153 |
| Sympathy | 122 |
| Help | 34 |
| Succour | 59 |
| News | 27 |
| Criticizing fake information | 10 |

*Table 2. Distribution of Bot-tweets within tweet Categories*

## 5　Discussion

Our preliminary findings revealed that 58 (0.2%) bots were found among the 20.000 power users – much less than previous studies estimated for Twitter communication (Varol et al., 2017). Thus, it can be assumed that during this particular crisis, social bots did not have a particular high influence on crisis communication regarding the *quantity* of the accounts. However, the results indicate that with 405 tweets in total and 7 tweets per account, the bot accounts were far more active than human accounts, which only posted 2 tweets per user on average. This result matches previous research stating that bots have an overall higher frequency (Davis et al., 2016). The results from our study indicate that this higher frequency may be also valid for benign bots. Taking into account that only a small sample was analysed, it will be necessary to see how the bot activity in the overall sample will turn out. One aspect that might be accountable for the low number of bot accounts is the timespan between the gathering of data (i.e. May 2017) and the date of the preliminary analysis (February 2018), which could lead to accounts being deleted from Twitter either for being bots or simply being deleted by their creators. This assumption is also backed by observations we made during the analysis of our dataset on Botometer via Python. During this process, we encountered instances which showed that accounts in the dataset could not be found on Twitter anymore.

The reason for the subsample of the 20,000 power users was to only choose the most-retweeted users. Since the 58 identified bots showed a high tweet activity, it is likely that there is much more bot activity within the remaining sample. This would also mean that those accounts are not retweeted as often as the 20,000 power users. However, focusing on the retweet count does not always show the most influential users (Mirbabaie et al., 2018). With an extensive Social Network Analysis, more influential bots may be identified. As our results indicate, many of the identified bots' messages fall into the category of *sympathy*. Hence, it might be possible that these accounts might also be able to provide emotional aid for people in need. Nevertheless, these findings also indicate that within the 20,000 most retweeted users, bot accounts are sparse yet active. Should this hold true for the overall sample it would also be an interesting finding for emergency agencies, who could incorporate the findings and in their twitter communication for crisis situations.

The analysis of the identified bots during a preliminary content analysis revealed that the content of all bots in our sample was classified as benign. This is contrary to findings in prior research. Most studies focused on the negative side of bots and their threats (Alarifi et al., 2016; Chu et al., 2012; Lokot and Diakopoulos, 2016) without assessing opportunities that could arise out of bots in general and in crisis situations more specific. One possibility would be that messages classified as benign were retweeted the most, because people could discriminate between those messages that really mattered and those that didn't, rendering a crisis situation an occasion, where the primary goal is to help people. In that regard, it might also be possible that malicious bots attempt to take advantage of the crisis in the aftermath of the attack. If the preliminary findings should hold true for the overall sample, it could also indicate that for emergency agencies, the use of social bots in crisis situations could be a way to be heard by the overall public and to efficiently spread important information. If during crisis situations benign messages by bot actors actually got retweeted the most (as indicated by our current findings from the top 20,000 retweeted users) this could hold potential for a professionalised coordination by emergency agencies. The introduction of social bots in this area could save a lot of time and provide a sophisticated coverage of news dissemination as well as emotional aid for the population.

## 6　Conclusion and Further Research

We could identify social bot activity in the researched sample of Twitter communication during the Manchester bombing. Initial findings suggest that the accounts in the current sample exclusively spread benign messages, having the potential to help people during an immediate crisis situation. Furthermore, the results indicate that overall, social bots have a higher tweet activity during crisis situations than human users. Thus, benign bots could exert a high influence on the sensemaking process on social media. An expansion of these findings is needed to get a clearer picture of the magnitude of the bot activity in the overall data sample. The preliminary findings presented in this paper indicate that bot activity is taking place during crisis situations and may even have the potential to actively help in a crisis situation. As our preliminary findings are based on a small sample of a rather active group of accounts, the logical next step will be to analyse bigger and multiple datasets. Besides the identification of social bots via the method already deployed on the current sample and an expansion of the content analysis, a social network analysis will be conducted to visualise and identify communities within the dataset. In that context, an inclusion of data regarding different influence measures such as the *Betweenness* or *Eigenvector* Centrality could show even more fruitful results with an increased number of influential bots.

Our preliminary findings have several limitations, which we plan to address in future research. First and foremost, the current findings are based on a sample of the actual dataset. While this sample consists of the 20,000 power users, the validity of the findings will be greatly enhanced by expanding the analysis to the whole dataset. Furthermore, the current analysis should be expanded, for one to include an expanded content analysis which's codebook could then also be used for further research on the field of bots in crisis situations. Also, a network analysis on the dataset would be valuable to gain better insights in the communication structure of bots during crisis. Based on an expansion of the analysis to the whole dataset it would also be interesting to compare the current case against other a) man made crises (such as other terror attacks) and b) natural disasters. As especially the former hold potential for the application of social bots, they are the best suited for initial analysis on the field of social bots in crisis communication. Besides the differentiation between different kinds of crises it is also important to look at the situation in different countries, and on different platforms. The current case also only focuses on the three days following the incident. Future research could expand this timeframe to gain further insights how the situation and the bot category distribution develops over time. Especially with regard to the possibility that the initial timespan may have been too short for creators of social bot accounts to intervene in the discussion. Further, a time series analysis should shed light on the dissemination of messages during the crisis and can show which actors are active at what time period. This step will provide insights into bot activity during certain time periods within a crisis and may for example show whether certain kind of bots (i.e. benign or malicious) are active during different time stamps and how dynamics shift over time.

In summary, the current paper presents preliminary findings on the activity of social bots during crisis communication. An extended version of this paper will build upon these preliminary findings and give extensive insights into the activity of social bots during crisis communication based on a large dataset of several million twitter accounts. Building upon these findings, IS researchers are able to tackle the phenomenon of participating bots in crisis situations and design guidelines as well as frameworks for the use of such data.